\documentclass[epj]{webofc}
\usepackage[varg]{txfonts}   

\usepackage{amsmath}

\woctitle{19th International Seminar on High Energy Physics QUARKS-2016}
\begin{document}
\title{Critical behavior of $U(n)$-$\chi^{4}$-model with antisymmetric tensor \\ order parameter coupled with magnetic field}

\author{N. V. Antonov\inst{1}\fnsep\thanks{\email{n.antonov@spbu.ru}} \and
        M. V. Kompaniets\inst{1}\fnsep\thanks{\email{m.kompaniets@spbu.ru}} \and
        N. M. Lebedev\inst{1}\fnsep\thanks{\email{nikita.m.lebedev@gmail.com}} \and
        }

\institute{St. Petersburg State University, Uljanovskaja 1, St. Petersburg, Petrodvorez, 198504 Russia}

\abstract{
  The critical behavior of $U(n)$-$\chi^{4}$-model with antisymmetric tensor order parameter at charged regime is studied by means of the field theoretic renormalization group (RG) at the leading order of $\varepsilon$-expansion (one-loop approximation). It is shown that RG equations have no infrared (IR) attractive charged fixed points. It is also shown that anomalous dimension of the order parameter in charged regime appears to be gauge dependent.
}

\maketitle

\section{Introduction}
\label{intro}
Investigation of critical behavior of Fermi particle systems is permanently in focus of attention due to the  problem of superconductive phase transition. The nature and properties of this phase transition was an open question for decades. There are two main approaches to the investigation of this problem. First one is based on microscopic description of quantum gas, and allows to perturbatively reveal phenomenon of superconductivity. Second one can be applied to the description of a small neighborhood of transition point and is based on Ginzburg–Landau effective Lagrangian studied by means of field theoretic renormalization group \cite{GreenBook}.

Because mean field theory can not be constructed in terms of Grassmann variables, the problem is to trace the connection between this two approaches. That connection has been traced in the work \cite{NKH}, where the authors have shown that critical modes of the simple microscopic model of fermions with $n$ possible spin projections and density-density interaction can be expressed in terms of antisymmetric tensor fields, whose means appears to be an order parameter. It was shown that in the vicinity of phase transition point the behavior of such system can be described by $\phi^4$-like theory, with two independent interaction terms. 

In the special case of $n=2$ and dimension $d=3$ that model coincide with the well known Ginzburg–Landau model \cite{LG},\cite{goodbook} and give same predictions. Also for $n=2$ and $d=4$ the model is formally close to the Higgs model \cite{Higgs}.

For the case of fermions with higher spins ($n>2$) resulting model was studied by the means of field theoretic renormalization group up to five-loops accuracy \cite{NKH}-\cite{NKK}. It was shown that RG equations of the model have no IR attractive fixed points, and corresponding RG flows (solutions of the RG equations for invariant couplings) always passes out from stability region of the model. Such situation is usually interpreted as a first order phase transition.

On the other hand it is expected that charged order parameter of superconductive phase transition should be also coupled with a magnetic field. Critical behavior of such model with $n$-component vector order parameter was studied in \cite{HLM}, \cite{DFM}, where it has been shown that charged fixed points at one-loop level exist only for sufficiently large $n$.

In this work we apply field theoretic RG to the $U(n)$-$\chi^{4}$-model with antisymmetric tensor order parameter coupled with a magnetic field. In the sections below we will show that charged fixed points exists only for $n>19$. At one loop-level they apper to be a saddle points, and lead to gauge dependent values of the critical exponent $\eta$.

\section{The model}
\label{model}

We study a model of a complex antisymmetric tensor field interacting with a magnetic field in $d-$dimensional Euclidean $\bf x$ space. Action functional of the model has the form:
\begin{equation}
S(\chi,{\bf A}) = tr((\nabla + ie_{0}{\bf A}_{0})\chi^{+}_{0} (\nabla - ie_{0}{\bf A}_{0})\chi_{0}) + \tau_{0} tr(\chi^{+}_{0} \chi_{0}) + \frac{g_{10}}{4}(tr(\chi_{0} \chi^{+}_{0}))^{2} + \frac{g_{20}}{4}tr(\chi_{0} \chi^{+}_{0} \chi_{0} \chi^{+}_{0}) + \nonumber
\end{equation}
\begin{equation}
+ \frac{1}{2}( \nabla \times {\bf A}_{0})^{2} + \frac{1}{2\alpha_{0}}(\nabla  {\bf A}_{0})^{2}.
\label{action}
\end{equation}
Here $\chi_{ik}$ and $\chi^{+ik}$ is an antisymmetric matrix fields (so that $\chi_{ik}=-\chi_{ki}$ and $\chi^{+ik}=-\chi^{+ki}$, $i,k=1,\ldots,n$), ${\bf A}_{0}$ is a magnetic field, $e_{0}$ is effective charge, $g_{10}$, $g_{20}$ are the coupling constants, $\tau_{0}$ is a deviation of temperature from the critical value and $\alpha_{0}$ is a gauge fixing parameter. Also here (and in analogous formulas below) integration over the $d-$dimensional $\bf x$ space is implied.

Note that field $\chi$ is twice covariant ($\chi \to \mathcal{U} \mathcal{U} \chi$), while field $\chi^{+}$ is twice contravariant ($\chi^{+} \to \chi^{+} \mathcal{U^{+}} \mathcal{U^{+}}$) with respect to transformations of the group $\mathcal{U} \in U(n)$.

Stability of the model require interaction part of the action to be positively defined. This requirement imposes following restriction on the coupling constants:
\begin{equation}
2g_{10} + g_{20} >0, \quad n g_{10} + g_{20} >0, \quad e^2 >0.
\label{stability}
\end{equation}

For $n=2$ and $n=3$ model is equivalent to $U(1)$ and $U(3)$ invariant $\phi^4$ models with a scalar and vector order parameter, respectively. Since analogues model of a vector order parameter has been studied at one-loop level in \cite{DFM} we have a way to check our results by comparing calculations in those particular cases.

\section{UV renormalization}
\label{UV renormalization}
In frameworks of dimensional regularization ultraviolet (UV) divergences in Green functions have the form of poles in $2\varepsilon = 4-d$ (the model is logarithmic for $d=4$). Their elimination requires renormalization of the model. Corresponding renormalized action can be obtained by multiplicative renormalization of fields and parameters of the model:
\begin{equation}
\tau_{0} = \tau Z_{\tau}; \quad \chi_{0} = \chi Z_{\chi}; \quad g_{0i} = g_{i}\mu^{2\varepsilon} Z_{g_{i}}; \quad {\bf A}_{0} = {\bf A} Z_{A}; \quad e_{0} = e\mu^{\varepsilon} Z_{e}; \quad \alpha_{0} = \alpha Z_{\alpha}.
\label{renormalization}
\end{equation}
Here $\tau$, $g_{i}$, $e$, $\alpha$ are renormalized analogues of bare parameters; $\chi$, ${\bf A}$ are renormalized fields and $\mu$ is a renormalization mass (arbitrary scale parameter). Also from the Ward identities follow relations:
\begin{equation}
Z_{e} = Z^{-1}_{a}; \ Z_{\alpha} = Z^{2}_{A}.
\label{ward}
\end{equation}
Then corresponding renormalized action has the form:
\begin{equation}
S_{R}(\chi,{\bf A}) = tr((\nabla + ie\mu^{\varepsilon}{\bf A})\chi^{+} (\nabla - ie\mu^{\varepsilon}{\bf A})\chi)Z^{2}_{\chi} + \tau tr(\chi^{+} \chi)Z_{\tau}Z^{2}_{\chi} + \frac{1}{2}(\nabla \times {\bf A})^{2}Z^{2}_{A} + \frac{1}{2\alpha}(\nabla  {\bf A})^{2} + \nonumber
\end{equation}
\begin{equation}
+ \frac{g_{1}\mu^{2\varepsilon}}{4}(tr(\chi \chi^{+}))^{2}Z^{4}_{\chi} Z_{g_{1}} + \frac{g_{2}\mu^{2\varepsilon}}{4}tr(\chi \chi^{+} \chi \chi^{+})Z^{4}_{\chi} Z_{g_{2}}
\end{equation}
We use minimal subtraction (MS) scheme where all renormalization constants have the form:
\begin{equation}
Z_{i}=1 + \sum_{p=1}^{\infty} A_{ip}(g_{1,2},e)\, \varepsilon^{-p}.
\label{ZMS}
\end{equation}
They are calculated from the 1-irreducible Green functions $\langle AA \rangle$, $\langle \chi^{+}\chi \rangle$, $\langle \chi^{+}\chi\chi^{+}\chi \rangle$. One-loop calculations gives:
\begin{equation}
Z^{2}_{A} = 1 - \frac{e^{2}n(n-1)}{6\varepsilon};\quad Z^{2}_{\chi} = 1 - \frac{e^{2}(3-\alpha)}{\varepsilon};
\label{Z_fields}
\end{equation}
\begin{equation}
Z^{2}_{\chi}Z_{g_{1}} = 1 + \frac{n^{2}-n+8}{4}g_{1} + (n-1)g_{2} + \frac{3}{4} \frac{g_{2}^{2}}{g_{1}}  - \frac{2e^{2}\alpha}{\varepsilon} + \frac{12e^{4}}{g_{1}\varepsilon};
\label{Z_g1}
\end{equation}
\begin{equation}
Z^{2}_{\chi}Z_{g_{2}} = 1 + 3g_{1} + \frac{2n-5}{4}g_{2} - \frac{2e^{2}\alpha}{\varepsilon}.
\label{Z_g2}
\end{equation}
Here and below we pass to rescaled couplings: $g_{1,2} = g_{1,2}/16\pi^{2}, \ e^{2} = e^{2}/16\pi^{2}$.

\section{RG functions}
\label{RG functions}
The RG functions of the model are defined by standard relations \cite{GreenBook}:
\begin{equation}
\gamma_{i}\equiv \widetilde {\cal D}_{\mu}\ln Z_{i}; \quad \beta_{i} \equiv \widetilde {\cal D}_{\mu} g_{i}; \quad \beta_{e} \equiv \widetilde {\cal D}_{\mu} e,
\label{rgf}
\end{equation}
where ${\cal D}_{x} \equiv x\partial_{x}$ for any variable x, and $\widetilde {\cal D}_{x}$ is the operation ${\cal D}_{x}$ at fixed bare parameters. In frameworks of MS scheme the RG functions can be expressed by the following simple expressions:

\begin{equation}
\gamma_{i} = - \left( 2{\cal D}_{g_{1}} + 2{\cal D}_{g_{2}} + {\cal D}_{e} \right) A_{i1}(g_{1,2},e);
\label{msg}
\end{equation}
\begin{equation}
\beta_{g_{i}} = g_{i}[-2\varepsilon-\gamma_{g_{i}}], \ i=1,2; \quad \beta_{e} = e[-\varepsilon-\gamma_{e}].
\label{msb}
\end{equation}
From equations (\ref{msg},\ref{msb}), explicit expressions (\ref{Z_fields}-\ref{Z_g2}) and Ward
identities (\ref{ward}) we obtain:

\begin{equation}
\gamma_{A} = -\gamma_{e} = \frac{1}{2}\gamma_{\alpha} = \frac{e^{2}n(n-1)}{6};
\end{equation}
\begin{equation}
\gamma_{\chi} = -e^{2}(3-\alpha);
\label{gamma_chi}
\end{equation}
\begin{equation}
\beta_{g_{1}} = -2\varepsilon g_{1} + \frac{n^{2}-n+8}{2} g_{1}^{2} + 2(n-1)g_{1}g_{2} + \frac{3}{2}g_{2}^{2} - 12e^{2}g_{1} + 24e^{4};
\label{beta_g1}
\end{equation}
\begin{equation}
\beta_{g_{2}} = -2\varepsilon g_{2} + 6g_{1}g_{2} + \frac{2n-5}{2} g_{2}^{2} - 12e^{2}g_{2};
\label{beta_g2}
\end{equation}
\begin{equation}
\beta_{e} = -e\varepsilon +\frac{e^{3}n(n-1)}{6}.
\label{beta_e}
\end{equation}

\section{Fixed points}
\label{Fixed points}

As it follows from RG equation \cite{GreenBook}, possible asymptotic regimes of the model are determined by asymptotic behavior of the system of equations for invariant coupling constants:
\begin{equation}
{\cal D}_s \bar g(s,g) = \beta (\bar g), \quad \bar g(1,g) = g.
\label{invg}
\end{equation}
Here $s=p/\mu$ is a non-dimensionalized momentum, $g=\{ g_{i},e \}$ is a full set of couplings and $\bar g(1,g)$ are the corresponding invariant variables. The IR ($s \to 0$) asymptotic of Green functions is determined by fixed points $g_{*}$ of system (\ref{invg}), which are determined by the requirement:
\begin{equation}
\beta (g_{*})=0.
\end{equation}
The type of fixed point is determined by eigenvalues of the matrix:
\begin{equation}
\omega = \partial\beta/\partial g |_{g=g_*}.
\label{omega}
\end{equation}
For IR attractive fixed point real part of all its eigenvalues is positive.

Analysis of expressions (\ref{beta_g1}-\ref{beta_e}) reveals two sets of fixed point corresponding to trivial ($e_{*}=0$) and nontrivial values of $e_{*}$. Both sets contain four fixed points. 

First one coincide with the set of fixed points of analogues model in absence of a magnetic field. It has been found and described in details in \cite{NKH}. This set consist of the following fixed points.
The trivial point:
\begin{equation}
g_{1}^{*} = 0; \quad g_{2}^{*} = 0,
\end{equation}
which is always IR repulsive. The point:
\begin{equation}
g_{1}^{*} = \frac{4 \varepsilon }{8-n+n^2}; \quad g_{2}^{*} = 0,
\end{equation}
which is IR attractive for $n=2$, and is a saddle point for any $n>2$. And two points with both nontrivial coordinates:
\begin{equation}
g_{1}^{*} = \frac{2 \left(77+8 n-4 n^2 \pm \sqrt{-(-5+2 n)^2 \left(-49-16 n+8 n^2\right)}\right) \epsilon }{392+151 n-19 n^2-24 n^3+4 n^4}; \nonumber
\end{equation}
\begin{equation}
g_{2}^{*} = \frac{4 \left(-70+103 n+5 n^2-24 n^3+4 n^4 \mp 6 \sqrt{-(-5+2 n)^2 \left(-49-16 n+8 n^2\right)}\right) \epsilon }{(-5+2 n) \left(392+151 n-19 n^2-24 n^3+4 n^4\right)},
\end{equation}
but they are real only for $n=2$. In this case they are either IR unstable or lie outside of physical region of parameters, which means that they can't be reached by RG flows.

Another set of fixed points is the original result, that we present by this article. It corresponds to the nontrivial value of $e_{*}$ equal to:
\begin{equation}
e^{2}_{*} = \frac{6\varepsilon}{n(n-1)},
\end{equation}
and consist of fixed points of two types.
Two points with coordinates:
\begin{equation}
g_{1}^{*} = \frac{2 \varepsilon \left(-4 n^6+16 n^5-87 n^4+286 n^3+2561 n^2 -2772 n \right) }{(n-1)^2 n^2 \left(4 n^4-24 n^3-19 n^2+151 n+392\right)} \pm \nonumber
\end{equation}
\begin{equation}
\pm \frac{\varepsilon\sqrt{-n^2 \left(2 n^2-7 n+5\right)^2 \left(8 n^6-32 n^5+2295 n^4-12014 n^3-265 n^2+48024 n+105840\right)}}{(n-1)^2 n^2 \left(4 n^4-24 n^3-19 n^2+151 n+392\right)};\nonumber
\end{equation}
\begin{equation}
g_{2}^{*} = \frac{4 \varepsilon \left(4 n^8-32 n^7+201 n^6-939 n^5+773 n^4+3771 n^3-6298 n^2 +2520 n \right) }{(n-1)^2 n^2 (2 n-5) \left(4 n^4-24 n^3-19 n^2+151 n+392\right)} \mp \nonumber
\end{equation}
\begin{equation}
\mp \frac{6 \sqrt{-n^2 \left(2 n^2-7 n+5\right)^2 \left(8 n^6-32 n^5+2295 n^4-12014 n^3-265 n^2+48024 n+105840\right)}}{(n-1)^2 n^2 \left(4 n^4-24 n^3-19 n^2+151 n+392\right)}.
\end{equation}
For any $n>1$ they have nontrivial imaginary part and therefore can not be reached by RG flows. Last two points have coordinates:
\begin{equation}
g_{1}^{*} = \frac{2 \left(n^4-2 n^3+37 n^2 -36 n \pm \sqrt{(n-1)^2 n^2 \left(n^4-2 n^3-359 n^2+360 n-2160\right)}\right) \epsilon }{(n-1)^2 n^2 \left(n^2-n+8\right)}; \nonumber
\end{equation}
\begin{equation}
g_{2}^{*} = 0;
\label{g*}
\end{equation}
They are real for $n>19$, but appear to be IR unstable (saddle) points with eigenvalues:
\begin{equation}
\omega_{1} = -\frac{2 \left(n^6-3 n^5+41 n^4-77 n^3+110 n^2 -72 n \mp 6 \sqrt{(n-1)^2 n^2 \left(n^4-2 n^3-359 n^2+360 n-2160\right)}\right) \varepsilon }{(n-1)^2 n^2 \left(n^2-n+8\right)};
\end{equation}
\begin{equation}
\omega_{2} = \pm \frac{2 \sqrt{(n-1)^2 n^2 \left(n^4-2 n^3-359 n^2+360 n-2160\right)} \varepsilon }{(n-1)^2 n^2};
\end{equation}
\begin{equation}
\omega_{e} = 2\varepsilon.
\end{equation}
Eigenvalue $\omega_{1}$ is always negative, while the sign of eigenvalue $\omega_{2}$ coincide with the sign in front of square root in (\ref{g*}).

\section{Conclusions}
\label{Conclusions}

We have shown that at one-loop level taking interaction with magnetic field into account lead to the appearance of four additional fixed points with nontrivial coordinate $e_*$, and two of them have real coordinates for $n>19$. Nevertheless both of them appear saddle points. It means that there is only one possible scenario of RG flows behavior: they will pass outside of stability region given by (\ref{stability}). Such situation is usually interpreted as a first-order phase transition. In turn it means that interaction with magnetic field do not change the qualitative picture of phase transition obtained in \cite{NKH}-\cite{NKK}.

Another interesting result is that anomalous dimension $\gamma_{\chi}$, and critical exponent $\eta = 2\gamma_{\chi}(g_*)$ as well, appears to be gauge dependent (\ref{gamma_chi}). It is not surprising because just the same situation takes place in case of the vector order parameter \cite{DFM}. However, $\eta$ become gauge dependent only at charged fixed point ($e^{2}_{*} \neq 0$). 
At the same time from renormalization of $\alpha$ follows that its RG flow should satisfy equation:
\begin{equation}
{\cal D}_s \bar \alpha = \bar \alpha \gamma_{\alpha},
\end{equation}
and should reach zero at charged fixed point. This means that gauge $\alpha = 0$ is the only gauge invariant with respect to renormalization (see \cite{DFM} for more detailed explanation).

\section*{Acknowledgments}

The authors are thankful to G.A. Kalagov and M.Yu. Nalimov for helpful discussions. N.V. Antonov is also thankful to R.Folk for the discussion of \cite{DFM}.

This work was supported by SPbSU grant 11.38.185.2014. N.M. Lebedev also thankful to RFBR project 16-32-00086 mol$\_$a, and to the "Dynasty" Foundation .

\end{document}